\documentclass[aps,prd,preprint,showpacs,amsmath,amssymb]{revtex4}
\usepackage{epsfig}
\usepackage{bm}

\newlength{\www}
\newcommand{\sopra}[2]{\settowidth{\www}{#1}#1\hspace{-\www}#2
\settowidth{\www}{#2}\hspace{-\www}\settowidth{\www}{#1}\hspace{\www}}
\newcommand{\barra}[1]{\sopra{#1}{/}}

\newcommand{\be}{\begin{equation}}
\newcommand{\ee}{\end{equation}}
\newcommand{\ba}{\begin{eqnarray}}
\newcommand{\ea}{\end{eqnarray}}

\newcommand{\thetabar}{\overline\theta}
\newcommand{\zetabar}{\overline\zeta}
\newcommand{\psibar}{\overline\psi}
\begin{document}


\newcommand{\PhysRevLett}[3]{Phys. Rev. Lett. {\bf #1}, #2 (#3)}
\newcommand{\PhysRevD}[3]{Phys. Rev. {\bf D #1}, #2 (#3)}
\newcommand{\NuclPhysB}[3]{Nucl. Phys. {\bf B #1}, #2 (#3)}
\newcommand{\PhysLettB}[3]{Phys. Lett. {\bf B #1}, #2 (#3)}
\newcommand{\ActaPhysPolB}[3]{Acta Phys. Pol. {\bf B #1}, #2 (#3)}


\title{\vspace{1cm}
Supersymmetric structure of electroweak Sudakov corrections
}

\author{Matteo Beccaria$^{a,b}$ and Edoardo Mirabella$^a$}

\affiliation{
$^a$Dipartimento di Fisica, Universit\`a di Lecce, Via Arnesano, 73100 Lecce, Italy.\\
$^b$ Istituto Nazionale di Fisica Nucleare, Sezione di Lecce
}

\begin{abstract}
Electroweak radiative corrections can be evaluated in the Sudakov approximation, a systematic high energy expansion
known to be relevant for the analysis of future collider experiments in the TeV energy range.
In the Minimal Supersymmetric Standard Model and at next-to-leading order,  Sudakov electroweak corrections satisfy 
remarkable relations at the one loop level.
Explicit computations in component fields are available for various different $2\to 2$ processes
relevant for Linear Collider or LHC physics. The Sudakov corrections turn out to be  equal or closely related
in several classes of processes differing by the replacement of certain final or initial states with their superpartners.
This fact suggests that supersymmetry is partially restored at high-energy. 
We analyze the supersymmetric structure of such relations by computing the Sudakov corrections 
in the framework of superfield perturbation theory.
As a simple application, we derive in full details an extended complete set of supersymmetric 
relations among different processes related by supersymmetry to the fundamental fermion pair 
production process $e^+e^-\to f\overline f$. 
\end{abstract}


\maketitle

\section{Introduction}
\label{Sec:Introduction}

The calculation of radiative corrections in models of fundamental interactions will play an
important role in the theoretical analysis of future collider experiments. In this paper, we work in the 
Minimal Supersymmetric Standard Model (MSSM), where higher order perturbative effects can be calculated systematically
due to renormalizability.

A general discussion of the structure of radiative corrections in the MSSM can be found in~\cite{Hollik1}. 
Actual calculations  involve the exchange of virtual states including, 
of course, the various supersymmetric partners of Standard Model particles.
In principle, supersymmetry (SUSY)  could simplify the analysis by providing an underlying structure.
In practice, supersymmetry is softly broken at low energy making very hard
to exploit the symmetry in the actual evaluation of definite physical processes~\cite{Martin}.

Thus, most of the existing calculations of radiative corrections in the MSSM
are performed in the familiar component formalism. Physical fields are treated separately and 
the SUSY multiplet structure is not fully exploited in the perturbative expansion.

This is an unpleasant situation, but the following considerations suggest the possibility of recovering supersymmetry 
to some extent in the phenomenological analysis of radiative corrections.
In the future collider experiments in the TeV energy range, 
in particular the Large Hadron Collider (LHC) and International Linear Collider (ILC),
the typical experimental situations will be characterized by 
large invariant masses. A high energy expansion seems therefore
a natural opportunity to achieve major simplifications in the structure of loop corrections.

Indeed, high energy expansions enjoy several useful features in this respect. 
First, helicity conservation reduce the number of relevant amplitudes. 
Second, it is known that radiative effects beyond production thresholds are rather smooth and can be described 
by the so-called logarithmic Sudakov expansion (LSE)~\cite{LSE}. 
In particular, in a MSSM scenario with light SUSY partners below the
typical value of 300-400 GeV, such expansions effectively apply at energies in the TeV range.
The practical consequences of LSE have been deeply analyzed in the Standard Model~\cite{Hollik2} and in the MSSM both at 
ILC~\cite{SudakovMSSM1,SudakovMSSM2,SudakovMSSM3} and at LHC~\cite{SudakovLHC}. 
Radiative corrections are expanded in a logarithmic variable that, in the electroweak case, is typically 
$\alpha_W \log(s/M_W^2)$, where $s$ is the squared center of mass energy and $\alpha_W = \alpha/\sin^2\theta_W$
with $\theta_W$ being the Weinberg weak mixing angle. Logarithms arise from ultraviolet and mass singularities.
At one loop and next-to-leading order (NLO), there appear contributions of the form $\alpha_W\log^2(s/M_W^2)$ and $\alpha_W\log(s/M_W^2)$  which 
have been shown to be sizable at the planned energies. For instance, at $\sqrt{s}=1$ TeV they give typical corrections of
tens per cent to the one loop cross sections.

The accuracy of LSE has been discussed by explicit comparisons with full one loop calculations of specific processes~\cite{SudakovMSSM2}. 
In practice, LSE leads to very simple and compact expressions that allow rather straightforward
phenomenological analysis~\cite{TanBeta}. With a conservative attitude,
processes where LSE predicts large observable effects have strong motivations for 
a full treatment.

We now come back to the issue of supersymmetry and consider what happens in the MSSM. Indeed, 
in this case we observe quite remarkable special features of LSE.
The most important is that LSE turns out to be independent from most of the
soft breaking terms at NLO. Therefore it isolates the dependence of observables on very specific parameters, like for instance
the vacuum alignment angle $\tan\beta$~\cite{TanBeta}.

Independence from soft breaking can be argued by  dimensional arguments in terms of the 
high energy mass suppression of low dimensional breaking operators. It suggests that some
amount of SUSY could be recovered at the level of LSE. This conjecture has  been checked
as a by-product of complete one loop calculations of various processes relevant for 
ILC and LHC~\cite{SudakovMSSM3} in the component approach.

The interesting result is that Sudakov corrections to processes involving particles 
in the same SUSY multiplets are closely related, as explained in full details below.
We shall denote briefly such kind of results Sudakov Supersymmetric Relations (SSR).
In this paper, we shall discuss a proof of such relations within the framework of superfield
perturbation theory.

The plan of the paper is the following. In Sec.~\ref{Sec:SSR} we discuss the specific details of SSR.
It will emerge that the Sudakov correction can be split into 
three contributions, so called universal, Yukawa and angular dependent.
In Sec.~\ref{Sec:Uni} we illustrate the supergraph calculation of the universal correction. 
In Sec.~\ref{Sec:Yukawa} we discuss the Yukawa correction. Finally, in Sec.~\ref{Sec:Box} we 
illustrate the evaluation of box supergraphs that compute the angular dependent correction.
In Sec.~\ref{Conclusions}, we summarize our results and discuss future developments.

\section{Sudakov Supersymmetric Relations}
\label{Sec:SSR}

We now review the observed SSR as they emerge from explicit one loop calculations in the MSSM in 
component fields. We  closely follow the notation of~\cite{SudakovMSSM3} and work at one loop.
We begin the discussion by considering the ILC process ($f\neq e$)
\be
\label{eq:basic}
e^+_\alpha e^-_\alpha\to f_\beta\overline{f}_\beta,
\ee
where $\alpha, \beta$ are chirality left ($L$) or right ($R$) indices. We denote by $s$ the squared center of mass (c.m.) energy
of the initial $e^+e^-$ pair. 

The electroweak Sudakov correction to the amplitude for the process (\ref{eq:basic})
is described at high energy and at next-to-leading logarithmic order in the following very simple form 
\be
\label{eq:LSEdetails}
A^{\rm Sudakov} = A^{\rm Born}(1+c_\alpha^{\rm U} + c_{\beta}^{\rm U} + c_\beta^{\rm Y} + c_{\alpha\beta}^{\rm ang}) + A^{\rm RG},
\ee
where the initial ($c_\alpha^{\rm U}$) and final  ($c_\beta^{\rm U}$)  universal coefficient $c^{\rm U}$ is 
\be
\label{eq:cuni}
c^{\rm U}_\alpha = \frac{1}{16\pi^2}\left(g^2 I_\alpha(I_\alpha+1)+g^{' 2}\frac{Y_\alpha^2}{4}\right)\left(2\log \frac{s}{M_V^2} - \log^2\frac{s}{M_V^2}\right),
\ee
where $I$, $T^3$ are the total and third component of isospin and $Y$ is the hypercharge. We denote by $g$ and $g'$ the 
SU(2) $\times$ U(1)$_{\rm Y}$ gauge coupling, following standard notation.

As we claimed, Eq.~(\ref{eq:LSEdetails}) is valid at NLO, {\em i.e.} when we neglect all terms which are not 
growing with the c.m. energy. 
The mass scale of the linear logarithm is not fixed at NLO in the logarithmic expansion.
Instead, the mass scale in the squared logarithm is not arbitrary. 
In components, several diagrams with exchange of gauge bosons do contribute. The mass  $M_V$  can be $M_\gamma$, $M_Z$ or $M_W$ 
according to the exchanged virtual gauge boson. For the photon, we set fictitiously $M_\gamma = M_Z$. This permits a unified treatment of the neutral current
electroweak interactions. In the application to a physical process, the correct procedure will require to 
reinsert $M_\gamma$ back into inclusive observables, include initial and final radiation effects, and finally take the 
physical $M_\gamma\to 0$ limit
(see for instance the detailed discussion about mass scale separation at two loop level in~\cite{Kuhn}).

The two terms in (\ref{eq:cuni}) have a clear separate origin from the neutral ($\gamma$, $Z$) and charged ($W^\pm$) current sectors, 
as shown by the Casimir operator of U(1)$_{\rm Y}$ or SU(2).
When the approximation $M_Z\simeq M_W$ is used, one recovers the above simple formula where the isospin and hypercharge contributions 
are collect together. If $M_Z$ and $M_W$ are kept distinct, the two contributions are separated accordingly~\cite{SudakovMSSM2}.

For simplicity, in this paper, we focus on the simpler neutral currents sector to illustrate the superfield technique. 
More precisely, we aim at recovering the SSR involving the hypercharge U(1)$_{\rm Y}$ part of the gauge group, {\em i.e.} for $c^{\rm U}$,
\be
c^{\rm U, neutral}_\alpha = \frac{1}{16\pi^2}\ g^{'\ 2}\frac{Y_\alpha^2}{4}\ \left(2\log \frac{s}{M_Z^2} - \log^2\frac{s}{M_Z^2}\right).
\ee
Intrinsic charged currents contributions can also be considered, but there are minor complications due to the triple gauge boson vertex.
The Sudakov correction is gauge invariant, and it is convenient to organize the various diagrams in a gauge invariant way
by means of the pinch technique~\cite{Pinch}. We leave details regarding the implementation of the pinch technique  with superfields to a separate work.

The Yukawa coefficient $c^{\rm Y}$ is present if $f$ is heavy and we keep its mass~\cite{TopLog}. Let us focus on the important 
case of top or bottom quark production. If we denote 
by $f'$ the SU(2) partner of $f$, we have 
\be
\label{eq:YukawaCorrection}
c^{\rm Y}_\beta = -\frac{g^2}{16\pi^2}\left(\frac{1+\delta_{\beta, R}}{2}\frac{\widehat{m}_f^2}{M_W^2} + \frac{\delta_{\beta, L}}{2}
\frac{\widehat{m}_{f'}^2}{M_W^2}\right)\log \frac{s}{M^2},
\ee
where $\widehat{m}_t = m_t/\sin\beta$ and $\widehat{m}_b = m_b/\cos\beta$, and $M$ is a mass scale not fixed at NLO.
It is  typically taken as $M = m_t$.
It is important to remark that it is only and precisely from this correction that a dependence on the MSSM parameters
arises at very high energy in this process.

The angular dependent correction is
\ba
c^{\rm ang}_{\alpha\beta} &=& -\frac{g^2}{16\pi^2}\log \frac{s}{M_V^2}\ 
\left[(\tan\theta_W^2 Y_\alpha Y_\beta+4 T^3_\alpha T^3_\beta)\ \log\frac{t}{u} + \right. \nonumber \\
&& \left. \frac{\delta_{\alpha, L}\delta_{\beta, L}}{1/4\ \tan\theta_W^2 Y_\alpha Y_\beta+T^3_\alpha T^3_\beta}
\left(\delta_{b,f}\ \log\frac{-t}{s}-\delta_{t,f}\ \log\frac{-u}{s}\right)\right] ,
\ea
where  $t$ and $u$ are standard Mandelstam variables. The hypercharge contribution is particularly simple 
\be
c^{\rm ang,\ neutral}_{RR} = -\frac{g^{' 2}}{4\pi^2}\frac{Y_\alpha}{2}\frac{Y_\beta}{2}\ \log\frac{s}{M_Z^2}\ \log\frac{t}{u} .
\ee
Finally, the Renormalization Group (RG) contribution $A^{\rm RG}$ contains the logarithms due to 
RG running of the coupling constants. This piece will be considered as {\em known} and not discussed in the following.
It can be found in details in~\cite{SudakovMSSM3}.

These expressions are remarkably simple and can be used as a starting point for the analysis of the phenomenological
consequences of radiative corrections, at least in the high energy regime. A detailed discussion of their accuracy, resummation properties, 
and relevance to the MSSM parameter space analysis can be found in the papers we have cited in the Introduction (see in particular the 
review~\cite{Hollik2}).

In this work, we shall be concerned with the SUSY structure of these corrections. Such a structure is illustrated by the following results.
Instead of (\ref{eq:basic}), we could consider the related process of sfermion production, again at ILC,
\be
e^+_\alpha e^-_\alpha\to \widetilde{f}_\beta\widetilde{f}_\beta^* .
\ee
Here, the index $\beta$ is not a chirality index but just denotes $\widetilde f_\beta$ as the sfermion partner of the chiral fermion $f_\beta$.
The remarkable result is that the Sudakov correction takes the same form as for fermion production
with the same coefficients, both for the universal, Yukawa, and 
angular dependent parts.

Also, if we are interested in charged Higgs production $e^+_\alpha e^-_\alpha\to H^+H^-$, we find again the same set
of corrections under the assignment $Q(H^-) = -1$, $T^3(H^-) = -1/2$. The only change is in the Yukawa term
that reads instead
\be
\label{eq:HiggsYukawa}
c^{\rm Y}_{H^-H^+} = -3\cdot \frac{g^2}{32\pi^2}\left(\frac{m_t^2}{M_W^2}\cot^2\beta + \frac{m_b^2}{M_W^2}\tan^2\beta \right)\log \frac{s}{M^2}.
\ee
Similar results are obtained for final charginos, neutralinos~\cite{TanBeta}, as well as for the initial states
which appear in LHC processes, {\em i.e.}, quarks, gluons~\cite{SudakovLHC}.
In the following, we shall denote by SSR (Sudakov SUSY relations) the equality of Sudakov corrections
for processes that differ by the replacement of some particle in the initial or final state with its superpartner.

The very existence of SSR proves that some amount of SUSY is recovered at the level of Sudakov correction, although supersymmetry is 
softly broken in the MSSM. As we said in the Introduction, we can argue that 
at high energy soft breaking terms are irrelevant and mass suppressed and SUSY relations are recovered.
An application of this idea can be found in~\cite{GBHC} where the structure
of gauge boson helicity conservation in the MSSM is analyzed under such hypothesis.

In principle, under the assumption of unbroken SUSY, some partial information can be derived from Ward identities.  
For instance, to match the corrections to the production of a fermion pair to those in the sfermion case, one could start from
the relation
\be
\label{eq:ward}
\delta_{\rm SUSY}\ \langle 0 | e^+_\alpha e^-_\alpha \widetilde{f}_\beta^* f_\beta | 0 \rangle = 0 .
\ee
The variation of the final state ($\widetilde{f}_\beta^* f_\beta$) connects the amplitudes for the two processes
$e^+_\alpha e^-_\alpha \to \widetilde{f}_\beta^* \widetilde{f}_\beta$ and $e^+_\alpha e^-_\alpha \to \overline{f}_\beta f_\beta$.
However, the terms coming from the variation of the initial state ($e^+_\alpha e^-_\alpha$) also contribute due to the possibility of exchanging a neutralino 
in the $s$ channel. Thus, Ward identities suggest that SSR can be extended to include the 
gaugino form factor $\lambda~\to~\varphi^*\psi$ in the equality between the Sudakov corrections to 
the vector form factors $V^\mu\to \psibar\psi$ and $V^\mu\to \varphi^*\varphi$. 
This will be immediately proved in the later calculations.
Beside these considerations, it must be checked in some details that soft terms are actually irrelevant and 
do not contribute at the Sudakov level to the right hand side of the variation Eq.~(\ref{eq:ward}).

It is clear that that to obtain more detailed results on the Sudakov logarithmic corrections
we should exploit in finer details their specific kinematical origin. Let us discuss the infrared contributions
that, for instance, lead to squared logarithms.
In spontaneously broken gauge theories the gauge boson masses $M_V$ are natural infrared regulators.
Thus, at large energy $E~\gg~M_V$, we can expect mass singularities to show up 
as powers of $\log(E/M_V)$, {\em i.e.} the infrared part of Sudakov corrections.
In the Standard Model, it is well known how to extract these (logarithmic) singular terms by a careful analysis
of the relevant diagrams~\cite{OriginalPapers}. Besides, the lowest order results can be 
improved by RG techniques in order to resum higher order effects and prove factorization properties.
Advanced examples of these resummation techniques include the treatment of angular dependent corrections whose numerical 
relevance in actual experiments is well known to be non negligible~\cite{SudakovMSSM1,Angular} and that, as we reviewed, also 
satisfy supersymmetric relations.
In principle, the extension to MSSM could be done in order to study the possible validity of SSR beyond one-loop.

Here, we begin with the one-loop contributions and  do not consider resummation issues. 
Instead, since we are interested in SSR, we show how the mass singularities can be isolated in a manifest supersymmetric way.
We stick to a detailed diagrammatic calculation because we believe that this is the clearest framework where 
soft breaking independence arises in a controlled and explicit way.
However, we do not work in components, but exploit instead superfield perturbation theory~\cite{Superspace,GSR}.

This technique is not much used in practical calculations
for phenomenological studies in the MSSM precisely because of the large amount of SUSY breaking. In principle,
soft terms can be included by the spurion technique~\cite{Spurion}. The method is somewhat cumbersome apart from 
the analysis of UV divergences or to the (related) construction of Renormalization 
Group evolution equations~\cite{RGevol}. For applications, component field calculations are certainly preferred.

Instead, Sudakov corrections can be computed safely and efficiently in the superfield approach, as this paper will show.
Indeed, the calculation is also largely simplified for the following reasons

\begin{enumerate}
\item Scheme independence: at one loop, Sudakov corrections are physical and renormalization scheme independent. Any scheme 
dependence drops out because scheme dependent subtractions do not introduce contributions growing with the energy. 
All the subtleties related to regularization consistency and to the need of restoring Ward identities~\cite{Hollik3}
do not show up at the next-to-leading logarithmic level.

\item Mass independence: we can insert fictitious masses in the superdiagrams, at least at NLO.
In all diagrams leading to Sudakov corrections, the mass terms are suppressed at high energy 
as a consequences of elementary properties of the basic loop integrals.
The only relevant mass scales entering the expressions for Sudakov 
corrections are the gauge boson masses unaffected by soft terms.
\end{enumerate}

We now consider the various supergraphs contributing the Sudakov corrections of various kinds. 
For each case, we compute the Sudakov expansion of the one-loop effective action in a manifest 
supersymmetric way and show how SSR are recovered quite easily.

\section{Universal Sudakov Correction}
\label{Sec:Uni}

\begin{figure}[htb]
\begin{center}
\hskip 0.1pt
\epsfig{file=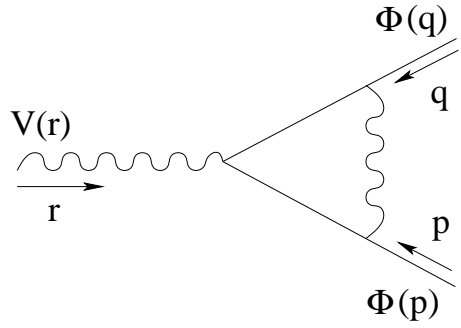,width=6cm}
\end{center}
\caption{Diagram giving the universal Sudakov correction to the $V\Phi\overline\Phi$ form factor.}
\label{VertexUniv}
\end{figure}

We begin with the $V\overline\Phi \Phi$ term in the one-loop effective action $\Gamma$. The supergraph
giving the Sudakov correction is shown in Fig.~(\ref{VertexUniv}) for the final 
state correction. Of course, the same (crossed) diagram give also the correction
to the initial state.
We want to retrieve the following result
\be
\Gamma_{V\overline\Phi\Phi}^{\rm Sudakov} = c^{\rm U}\ \Gamma_{V\overline\Phi\Phi}^{\rm tree},
\ee
where $c^{\rm U}$ is given in Eq.(\ref{eq:cuni}). After expansion of the superfields $V$, $\Phi$, and $\overline\Phi$,
this proves SSR for the gauge universal Sudakov correction. 

The value of the supergraph is (see for instance the analogous computation in 
$O(4)$ extended super Yang-Mills~\cite{GSR}):
\ba
\label{eq:Vertex}
\Gamma_{V\overline\Phi\Phi} &=& -\frac{ig_0^2}{2}\int \frac{d^4k}{(2\pi)^4}\frac{1}{D_1 D_2 D_3} \\
&&
\int d^4\theta\ \Phi(p, \theta)\ \overline\Phi(q, \theta)
\left[(k+q)^2-\frac 1 2 D (\barra{k}+\barra{q})\overline{D}-\frac{1}{16} D^2 \overline{D}^2\right]
V(r, \theta), \nonumber
\ea
where $g_0$ is the neutral $V$ coupling and $D_1 = k^2-M_Z^2$, $D_2 = (k+q)^2-M_\Phi^2$, $D_3 = (k-p)^2-M_\Phi^2$.
The convention for chiral and vector superfields is 
\ba
\Phi &=& \varphi + \theta\psi + \theta^2 F -i\theta\sigma^\mu\thetabar\ \partial_\mu\varphi -\frac{i}{2}\theta^2\thetabar\overline\sigma^\mu\ \partial_\mu\psi
-\frac{1}{4}\Box \varphi\ \theta^2\thetabar^2 , \\
V_{WZ} &=& \theta\sigma^\mu\thetabar\ V_\mu -i\theta^2\thetabar\overline\lambda+i\thetabar^2\theta\lambda+\frac 1 2 D\ \theta^2\thetabar^2 ,
\ea
where, for simplicity, we have written $V$ in the Wess-Zumino gauge. This is possible for external $V$ fields in gauge invariant
amplitudes. Instead, virtual $V$ exchange will involve all the unphysical degrees of freedom of $V$.
The covariant derivatives $D$, $\overline D$ acting on $V$ are conveniently written after contraction with auxiliary spinors $\zeta$, $\zetabar$
and read ($\barra{r} \equiv \sigma^\mu r_\mu$)
\be
\zeta D = \zeta\partial+\zeta\barra{r}\thetabar,\qquad \zetabar\overline{D} = -\overline{\partial}\zetabar-\theta\barra{r}\zetabar .
\ee

We now analyze the three terms inside the square brackets in $\Gamma_{V\overline\Phi\Phi}$. 
The first term is proportional to the Passarino-Veltman (PV) function $B_0$ 
evaluated at one of the external $\Phi$ squared momenta. 
It is divergent and must be regulated but does not give Sudakov logarithms being independent from the c.m. energy.
Basic definitions and asymptotic expansions of the PV functions
can be found in full generality in~\cite{Loop}. 

In the second term of $\Gamma_{V\overline\Phi\Phi}$ we require the finite loop integral (in the large $r^2 \gg M_i^2, p^2, q^2$ regime)
\be
\label{eq:loopuniversal}
\int\left(\frac{dk}{2\pi}\right)^4\frac{(k+q)_\mu}{(k^2-M_V^2)[(k+q)^2-M^2_\Phi][(k+q+r)^2-M^2_\Phi]} = -\frac{i}{16\pi^2}(C_{11}q + C_{12}p)_\mu .
\ee
The PV functions $C_{11}$ and $C_{12}$  have the following asymptotic expansion where $\sim$ denotes equality modulo terms not growing with $r^2$
\ba
C_{11} &\sim& \frac{1}{2r^2}\left(2\log\frac{r^2}{M_V^2}-\log^2\frac{r^2}{M_V^2} \right), \\
C_{12} &\sim& -\frac{1}{r^2}\ \log\frac{r^2}{M_V^2}   .
\ea
The third term can be integrated by parts moving the $D$ and $\overline D$ operators on the chiral and antichiral
fields $\Phi$, $\overline\Phi$. By chirality, we have $D_\alpha\overline\Phi = D_{\dot\alpha}\Phi = 0$. Also, 
the expressions $D^2\Phi$ and $\overline{D}^2\overline\Phi$ are not vanishing but are mass suppressed as a 
consequence of the equations of motion of $\Phi$, $\overline\Phi$. To see this, we contract the effective action 
vertex $V\Phi\overline\Phi$ with the propagator
\be
\label{eq:prop}
\langle 0 | T\left\{ \Phi(p, \theta_1, \thetabar_1)
\ \overline\Phi(-p, \theta_2, \thetabar_2)\right\} |0\rangle = \frac{1}{16}\frac{i}{p^2-M_\Phi^2}\ \overline D_1^2 D_1^2\delta_{12},
\ee 
where $\delta_{12} = (\theta_1-\theta_2)^2(\thetabar_1-\thetabar_2)^2$. If we now act with $D^2$ and exploit $D^2\overline{D}^2 D^2 = 16 \Box D^2$,
we see that, apart from mass corrections, any effective vertex with a $D^2\Phi$ or $\overline D^2\overline\Phi$ term has not
the necessary pole to survive the on-shell projection.

The Sudakov correction comes therefore entirely from the second term 
\be
\Gamma^{\rm Sudakov}_{V\overline\Phi\Phi}(q, p, r) = 
\frac{ig_0^2}{4}\int\frac{d^4k}{(2\pi)^4}\int d^4\theta\ \overline\Phi(q)\Phi(p)\ (k+q)_{\alpha\dot\alpha}D^\alpha
\overline D^{\dot\alpha} V(r) .
\ee
We perform the loop integral and replace $k+q\to (-i/16\pi^2)(C_{11}q + C_{12}p)$. Integrating by parts the $D$, $\overline D$
operators we find 
\ba
\label{eq:tmp1}
\lefteqn{\Gamma^{\rm Sudakov}_{V\overline\Phi\Phi}(q, p, r) = } && \\
&=& \frac{g_0^2}{4\cdot 16\pi^2} \int d^4\theta\ 
\left[\overline D^{\dot\alpha}\overline\Phi(q)\ D^\alpha\Phi(p) + \overline\Phi(q)\overline D^{\dot \alpha}D^\alpha\Phi(p)\right]\ 
(C_{11}\barra{q}+C_{12}\barra{p})_{\alpha\dot \alpha} V(r) . \nonumber
\ea
Since we are interested in the Sudakov enhanced terms, we can further simplify. Indeed, we observe that 
the combination $(\barra{p})^{\dot\alpha \alpha}\partial_\mu D_\alpha \Phi$ and the similar one for $\overline\Phi$
are mass suppressed as follows again from Eq.~(\ref{eq:prop}). The first term of Eq.~(\ref{eq:tmp1})
does not contribute. Anticommuting $D$ and $\overline D$ in the second term and taking the trace we obtain
\be
\label{eq:clog}
\Gamma^{\rm Sudakov}_{V\overline\Phi\Phi}(q, p, r) = 
\frac{g_0^2}{16\pi^2} (q\cdot p)\ C_{11} \int d^4\theta\ 
\overline\Phi(q)\ \Phi(p)\ V(r).
\ee
We can now replace $C_{11}$ by its asymptotic expansion and identify $g_0 = 2 (g' Y/2)$ where $g' Y/2$ is the U(1)$_{\rm Y}$
standard coupling. A factor 2 is needed to conform to standard normalization of the gauge field in components.
The final result is
\be
\label{eq:ResultUniversal}
\Gamma^{\rm Sudakov}_{V\overline\Phi\Phi} = \frac{1}{16\pi^2} \left(g' \ \frac{Y}{2}\right)^2 \ 
\left(2\log \frac{r^2}{M_Z^2} - \log^2\frac{r^2}{M_Z^2}\right)
\int d^4\theta\ \overline\Phi\ \Phi\ V = c^{\rm U}\ \Gamma^{\rm tree}_{V\overline\Phi\Phi}.
\ee
We have obtained in a very straightforward way the SSR for the neutral part of the correction $c^{\rm U}$. The scale $M_V$ has been set to $M_Z$
since the possible soft corrections to the gaugino mass cannot affect the squared logarithm scale.
As a technical comment, the unphysical scalar component of $V$ does propagate along the internal $V$
line, but it can be checked that it does not give Sudakov logarithms. Thus, the result in Eq.~(\ref{eq:ResultUniversal})
is the correct physical gauge invariant one.

To summarize, the superfield calculation of the effective action has taken into account simultaneously
the correction to the three specific vertices appearing in the expansion of 
$\int d^4\theta\ V_{WZ}\overline\Phi\Phi$, {\em i.e.}
\be
\label{eq:menu}
\psi \sigma^\mu \psibar\ V_\mu, \qquad i(\varphi\partial^\mu\varphi^*-\partial^\mu\varphi \varphi^*) V_\mu,
\qquad \overline\lambda\ \psibar \varphi + \mbox{h.c.} .
\ee
The first two appear in the initial and final universal correction to the amplitudes for producing 
a $f\overline{f}$ or $\widetilde{f}\widetilde{f}^*$ pair. They have been proved to be equal in agreement with SSR. 
The third one is a byproduct of the superfield calculation and, by supersymmetry, gets an identical correction.
Hence, the SSR can be extended in supersymmetric way to the gaugino form factor $\lambda\to \phi^*\psi$ 
in full consistency with the Ward identity (\ref{eq:ward}). As a check, an explicit calculation in components
confirms this result.

\section{Yukawa Sudakov Correction}
\label{Sec:Yukawa}

The Yukawa corrections proportional to the squared heavy quark masses arise from a different kind of supergraph
and  have special features that we now discuss in details.
We start from the corrections $\sim m_t^2$. Later, we shall include those $\sim m_b^2$. 
The Sudakov Yukawa corrections can be treated
in a simplified and independent sector of the MSSM governed by the Lagrangian
\be
L = \sum_{\Phi} \int d^4\theta\ \overline{\Phi}\ e^{g_0V}\ \Phi + y_t \varepsilon^{\alpha\beta} \int d^2\theta\ 
(\overline u \ Q_\alpha H_{u\ \beta} + {\rm h.c.}),
\ee
where the vector $V$ is external and, in the MSSM, stands for one of the neutral gauge bosons $Z$, $\gamma$. The coupling $y_t$ is the relevant Yukawa 
coupling proportional to $m_t$ and reads at lowest order 
\be
y_t = \frac{g\ m_t}{\sqrt{2} M_W \sin\beta} ,
\ee
where $g$ is the SU(2) gauge coupling. The various chiral fields are as follows. 
The SU(2) singlet $\overline u$ contains as physical fields
the right component of top $t_R$ and its SUSY partner $\widetilde{t}_R$. The SU(2) doublet $Q$
contains two chiral fields $u$, $d$, whose components are the left handed top and bottom, with their superpartners. 
The SU(2) doublet $H_u$ is one of the two Higgs doublets of the MSSM and contains two chiral fields $H^+$, $H^0$
\be
Q = \left(\begin{array}{cc}u \\ d\end{array}\right),\qquad
H_u = \left(\begin{array}{cc}H^+_u \\ H^0_u\end{array}\right).
\ee
Since $V$ is external, the calculation can be entirely done in a gaugeless limit of the MSSM, 
precisely as in \cite{Barbieri} where  the subleading correction ${\cal O}(G_\mu m_t^2)$ 
to the $Z\to b\overline{b}$ vertex and $\rho$ parameter is computed in a gaugeless limit of the Standard Model.

\begin{figure}[htb]
\begin{center}
\hskip 0.1pt
\epsfig{file=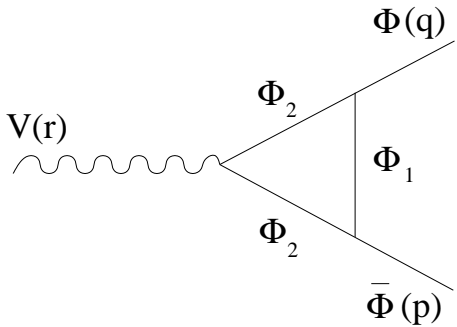,width=6cm}
\end{center}
\caption{Diagram contributing the $y_t^2$ Yukawa terms}
\label{VertexYukawa1}
\end{figure}

The Sudakov correction comes from the diagram  proportional to $g\ y_t^2$ shown in Fig.~(\ref{VertexYukawa1}) 
This is similar to the previous  $\Gamma_{V\overline\Phi\Phi}$, but differs in the chiral structure because of the exchange 
of a $\Phi$ field, instead of a vector $V$. The D-algebra details can be worked out and 
the expressions we derived for the universal correction are still valid apart from 
trivial changes in the couplings and, most important, the swap $C_{11}\leftrightarrow C_{21}$.
This will be the technical reason for the correction being linear and not quadratic in the logarithm of the 
center of mass energy.

To perform the actual calculation, we should evaluate a few diagrams according to specific final state and exchanged virtual $\Phi$ fields. 
Indeed, due to the explicit SU(2) structure of the Higgs field it is not possible to 
join together the photon and the $Z$. 
However, instead of considering the separate corrections to the $\gamma$ and $Z$ vertices it is more instructive to 
relate the single logarithm appearing in $c^{\rm Y}$ to the external $\Phi$ self energy divergences. 
This is possible due to the following  remarks.
\begin{enumerate}
\item The linear logarithm and the divergence $\Delta$ of the vertex proportional to $y_t^2$
appear in the fixed combination $\Delta-\log(s/M^2)$. This follows from the actual expression of the diagram 
in Fig.~(\ref{VertexYukawa1}). It is enough to substitute $C_{11}\to C_{12}$ in Eq.~(\ref{eq:clog})
and to combine the logarithm coming from the asymptotic expansion of $C_{12}$ with the divergence of the $B_0$ function from the first term of 
Eq.~(\ref{eq:Vertex}).

\item The divergence in the vertex is canceled by the external $\Phi$ self energies, also proportional to $y_t^2$.
This follows from the fact that the divergence of the sum of diagrams shown in Fig.~(\ref{VertexYukawa3})
and dictated by the SU(2) invariant  $\varepsilon$ tensor 
is proportional to the sum of charges of the three chiral fields $\Phi_{1,2,3}$.
Thus, cancellation
results from the Yukawa interaction being a neutral SU(2) singlet, as it must. 
\begin{figure}[htb]
\begin{center}
\hskip 0.1pt
\epsfig{file=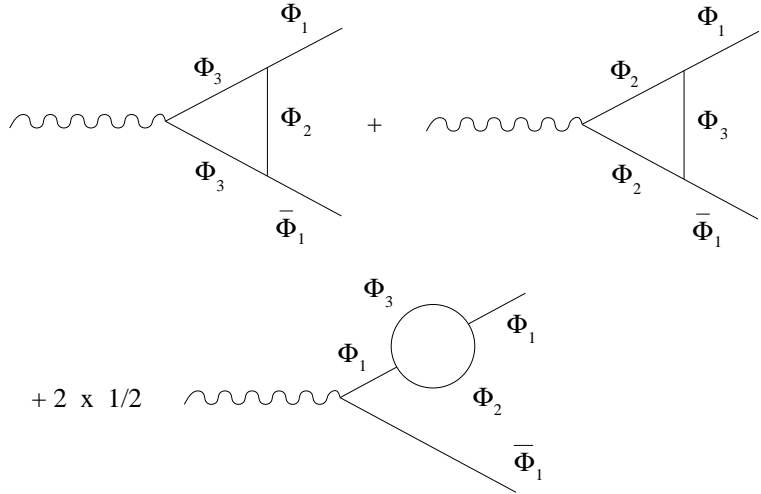,width=10cm}
\end{center}
\caption{Cancellation of the divergence in the terms proportional to $y_t^2$. It is given by the sum of the above
Yukawa diagrams where $\Phi_{1,2,3}$ can be $H_u$, $Q$, or $\overline{u}$ as allowed by the 
SU(2) invariant coupling $\overline{u}\ Q\cdot H_u$.}
\label{VertexYukawa3}
\end{figure}
\end{enumerate}

The above observations link the calculation of the logarithmic Yukawa correction to 
the external propagators which are are quite simple and independent from the off-shell vector $V$. 
Indeed, the coefficient of the logarithm in the vertex must be equal to the coefficient of the divergence in the 
self-energy.
The correction to the kinetic term $\int d^4\theta\ \Phi\overline\Phi$ in the effective action is well known and reads
\be
\Gamma(\Phi, \overline\Phi) =  y_t^2\ \int\frac{d^4p}{(2\pi)^4}\ A(p)\ d^4\theta\ \overline\Phi(-p, \theta)\ \Phi(p, \theta),
\ee
where 
\be
A(p) = \frac{1}{16\pi^2}\Delta + \mbox{finite}.
\ee 
In addition, counting the possible diagrams and  the color factors, we have the following multiplicity factors
for the superfield propagators
\ba
\Gamma(\overline u, \overline u^{\ *})  &:& 2, \\
\Gamma(Q_\alpha, \overline Q_\beta) &:& \delta_{\alpha\beta}, \\
\Gamma(H_{u\ \alpha}, \overline H_{u\ \beta}) &:& N_c\ \delta_{\alpha\beta}, \ \ (N_c = 3).
\ea
In conclusion, the relative Yukawa correction to the various processes one can consider 
is obtained in terms of the common coefficient
\be
c = -\frac{g^2}{2\cdot 16\pi^2}\frac{\widehat{m}_t^2}{M^2_W}\ \log\frac{s}{m_t^2} = -\frac{\alpha}{8\pi s_W^2}\frac{\widehat{m}_t^2}{M^2_W}\ \log\frac{s}{m_t^2} ,
\ee
as in the following list
\ba
c^{\rm Y}(u_R, \overline u_R) &=& c^{\rm Y}(\widetilde u_R, \widetilde u_R^*)  = 2c, \\
c^{\rm Y}(u_L, \overline u_L) &=& c^{\rm Y}(\widetilde u_L, \widetilde u_L^*)  = c, \\
c^{\rm Y}(d_L, \overline d_L) &=& c^{\rm Y}(\widetilde d_L, \widetilde d_L^*)  = c, \\
c^{\rm Y}(H_u^+, H_u^-) &=& c^{\rm Y}(\widetilde H_u^+, \widetilde H_u^-) = 3c.
\ea
If we now add the $y_b^2$ term $-y_b\ \overline{d}\ Q\cdot H_d$ (where the second Higgs doublet is $H_d = (H_d^0, H_d^-)^T$)
we recover the full Yukawa correction Eq.~(\ref{eq:YukawaCorrection}) for final fermions or sfermions. 
In particular, we proved SSR for this kind of correction with almost no calculations.

For final charged Higgs bosons, we must recall that the physical $H^+$ field is 
defined as the combination $H^+ = \cos\beta H_u^+ + \sin\beta H_d^-$. Therefore, for $H^+H^-$ production, we have to scale
$m_t\to m_t \cos^2\beta$ and $m_b\to m_b \sin^2\beta$ in the above expressions obtaining precisely Eq.~(\ref{eq:HiggsYukawa}).

\section{Angular dependent Sudakov Correction}
\label{Sec:Box}

We now turn to the final piece of the calculation, the angular corrections coming from gauge boson exchange in boxes.
The effective four point vertex is proportional to $\int d^4\theta \ \Phi_1\Phi_2\overline\Phi_3\overline\Phi_4$,
where $\Phi_i$ are the chiral and antichiral fields required to describe the initial and final states of the 
scattering process.
For instance, in four fermion scattering at tree level, the sum of $\gamma$ and $Z$ exchange leads to the Dirac structure 
$\gamma^\mu\otimes \gamma_\mu$. For right handed fermions in two spinor notation, this is proportional to 
$\sigma^\mu\otimes\sigma_\mu$ that can written in the form  $(\overline\psi_1\overline\psi_4)\ (\psi_2\psi_3)$
by a Fierz transformation. This is precisely the fully fermionic component of the above  superspace integral.

\begin{figure}[htb]
\begin{center}
\hskip 0.1pt
\epsfig{file=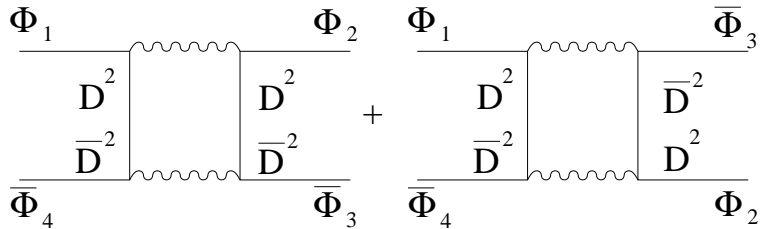,width=10cm}
\end{center}
\caption{Direct and crossed box diagrams giving the angular Sudakov correction.}
\label{Box1}
\end{figure}

The angular dependent Sudakov correction comes from the two diagrams drawn in Fig.~(\ref{Box1}) with their
detailed chiral structure and from the two diagrams involving the $V^2\Phi\overline\Phi$ vertex  shown
in Fig.~(\ref{Box1}). 
\begin{figure}[htb]
\begin{center}
\hskip 0.1pt
\epsfig{file=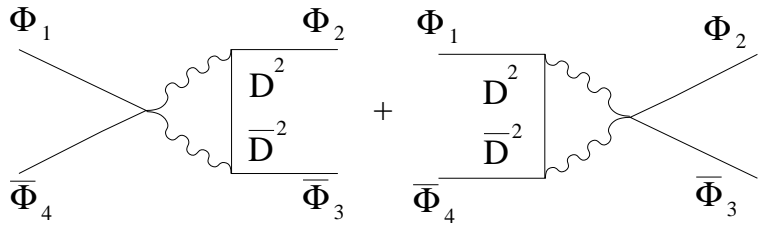,width=10cm}
\end{center}
\caption{Box diagrams with one four leg vertex $V^2\Phi\overline\Phi$.}
\label{Box1bis}
\end{figure}
For simplicity, we are assuming that the initial and final states do not belong to the same (lepton or quark) family.

For details about D-algebra manipulations, we refer to~\cite{Superspace,GSR}. We just illustrate the general algorithm
at one loop
since box calculations with superfields are not usual.
The various $D^2$ and $\overline{D}^2$ operators are moved away from their position by integration by parts
in $\theta$ space. They can be transferred to other internal lines or to external fields. When all the $D^2$ and $\overline{D}^2$ 
on internal lines are on the same line, we obtain a zero result if there are less then four factors. If we have precisely 
$D^2\overline{D}^2$, we can simply remove them. 
In particular, the D-algebra of Fig.~(\ref{Box1bis}) is trivial.
If we have more than four $D$, $\overline D$ factors, we can exploit the D-algebra 
$\{D_\alpha, \overline D_{\dot\alpha}\} = 2i\sigma^\mu_{\alpha \dot\alpha}\ \partial_\mu$ to reduce their number. Useful relations 
are $D_\alpha D_\beta D_\gamma = 0$, $D^2\overline D^2 D^2 = 16 \Box D^2$ and so on. 
After doing the D-algebra, we are left with a single $d^4\theta$ integration and a standard loop integral.

\begin{figure}[htb]
\begin{center}
\epsfig{file=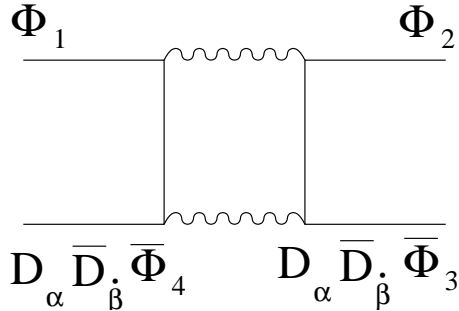,width=6cm}
\end{center}
\caption{Direct Box diagram: surviving term after D-algebra manipulations and neglecting 
mass suppressed terms.}
\label{Box2}
\end{figure}

The direct box diagram is reduced to Fig.~(\ref{Box2}).
As in the calculation of the  universal correction, we neglect systematically all mass suppressed contributions $D^2\Phi$, 
$\overline D^2\overline \Phi$.
Also, by definition $D\overline\Phi = \overline{D}\Phi = 0$. Therefore, we can simplify 
$D_\alpha \overline D_{\dot \alpha}\overline \Phi(p) \sim 2 p_{\alpha\dot\alpha}\overline\Phi(p)$.
The diagram is thus proportional to $2p_3\cdot p_4\ D_0$ where $D_0$ is the four point basic PV function
defined by 
\be
\int\frac{d^4k}{i\pi^2}\frac{1}{D_1 D_2 D_3 D_4} = D_0,
\ee
with
\be
D_1 = k^2-M^2, D_2 = (k+p_1)^2-M^2,
\ee
\be
D_3 = (k+p_1+p_2)^2-M^2, D_4 = (k+p_1+p_2+p_3)^2-M^2.
\ee
For simplicity and somewhat heuristically, we  put the same mass $M$ in all propagators. The final result will be 
independent from $M$ at next-to-leading logarithmic level. We could of course, keep separate masses inside
the loop and find the same result. We define the Mandelstam variables as $s = (p_1+p_4)^2$, $t = (p_1+p_2)^2$, $u = (p_1+p_3)^2$. 
In the large $s$ limit, with fixed ratios $t/s$, $u/s$, the asymptotic expansion of the $D_0$ function is 
\be
D_0\sim \frac{1}{st}\left(\log^2\frac{s}{M_V^2} + \log^2\frac{t}{M_V^2}\right).
\ee
The relative value of the one loop correction with respect to the tree scattering amplitude is then
\be
c^{\rm ang}_{\rm direct} = -\frac{g_0^2}{32\pi^2}\left(\log^2\frac{s}{M_V^2} + \log^2\frac{t}{M_V^2}\right).
\ee 

\begin{figure}[htb]
\begin{center}
\hskip 0.1pt
\epsfig{file=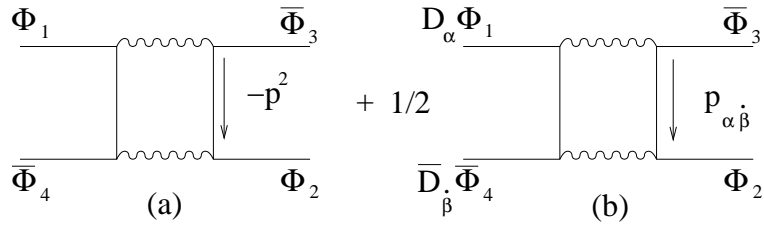,width=10cm}
\end{center}
\caption{Crossed Box diagram: surviving contributions after D-algebra manipulations and neglecting mass
suppressed terms.}
\label{Box3}
\end{figure}

To evaluate the crossed diagram it is convenient to perform D-algebra making all the $D$, $\overline D$ operators act on 
the fields $\Phi_1$ and $\overline\Phi_4$. This leads to the combination shown in Fig.~(\ref{Box3}).
Apart from trivial normalizations and couplings that we shall insert in the end, the diagram (a) gives a $C_0$
function~\cite{Loop} due to the canceled propagator between fields 2 and 3. The same is true for the two diagrams in Fig.~(\ref{Box1bis}).
The $C_0$ function and its expansion for $s=(p_1+p_2^2) \gg p_i^2, m_i^2$ are 
\be
C_0(p_1, p_2) = \int\frac{d^4k}{i\pi^2} \frac{1}{(k^2-m_1^2)((k+p_1)^2-m_2^2)((k+p_1+p_2)-m_3^2)} \sim \frac{1}{2s}\log^2\frac{s}{m_2^2} .
\ee
Keeping only the terms growing with energy we obtain
\be
(a) + \mbox{Fig.~(\ref{Box1bis})}= \frac{1}{2s}\log^2\frac{s}{M^2} .
\ee
Diagram (b) requires the expansion of the $D_{1i}$ functions defined by
\be
\int\frac{d^4k}{i\pi^2}\frac{k^\mu}{D_1 D_2 D_3 D_4} = \sum_{i=1}^3 D_{1i} p_i^\mu .
\ee
These are given asymptotically by 
\ba
D_{11} &\sim & \frac{1}{2st}\left(-\log^2\frac{s}{M^2}-2\log^2\frac{t}{M^2}\right), \\
D_{12} &\sim & \frac{1}{2st}\left(-\log^2\frac{s}{M^2}-\log^2\frac{t}{M^2}\right), \\
D_{13} &\sim & -\frac{1}{2st}\log^2\frac{t}{M^2} .
\ea
As a final step, we must evaluate
\be
\label{eq:boxtmp}
(b) = \frac{1}{2}\int d^4\theta\ D_\alpha\Phi_1\ \Phi_2\ \overline\Phi_3\ \overline{D}_{\dot\alpha}\overline\Phi_4
(\overline\sigma^\mu)^{\dot\alpha \alpha}(\widetilde{D}_{11}\ p_1 + \widetilde{D}_{12}\ p_3 + D_{13}\ p_2)_\mu ,
\ee
where $\widetilde{D}_{1i} = D_{1i}+D_0$. We already remarked that 
$(\barra{p})^{\dot\alpha \alpha} D_\alpha \Phi(p)$ and its conjugate are mass suppressed. 
Thus, we manipulate each of the terms in Eq.~(\ref{eq:boxtmp}) in order to have a $D$ (or $\overline D$)
operator on each derived field ({\em i.e.} fields accompanied by a factor of the associated momentum).

The term $\sim p_1$ is already in this form and does not contribute. The $\sim p_3$ term
is treated as follows
\ba
\lefteqn{
\int d^4\theta\ D_\alpha\Phi_1\ \Phi_2\ \overline\Phi_3\ \overline{D}_{\dot\alpha}\overline\Phi_4 (\overline{\barra{p}}_3)^{\dot\alpha \alpha} = } && \\
&=& -\int d^4\theta\ \overline{D}_{\dot\alpha}D_\alpha\Phi_1\ \Phi_2\ \overline\Phi_3\ \overline\Phi_4 (\overline{\barra{p}}_3)^{\dot\alpha \alpha} +\dots =  
4 p_1\cdot p_3 \int d^4\theta\ \Phi_1\ \Phi_2\ \overline\Phi_3\ \overline\Phi_4 , \nonumber
\ea
where we used integration by parts of the $D$, $\overline D$ operators, because of momentum conservation.
In a similar way we obtain for the $\sim p_2$ term
\be
\int d^4\theta\ D_\alpha\Phi_1\ \Phi_2\ \overline\Phi_3\ \overline{D}_{\dot\alpha}\overline\Phi_4 (\overline{\barra{p}}_2)^{\dot\alpha \alpha} = 
 -4 p_2\cdot p_4 \int d^4\theta\ \Phi_1\ \Phi_2\ \overline\Phi_3\ \overline\Phi_4 .
\ee
Hence, 
\be
(b) = \frac{1}{2}\ 2u (\widetilde{D}_{12}-D_{13}) = \frac{1}{2s}\left(\log^2\frac{s}{M^2} + 2\log^2\frac{u}{M^2} \right).
\ee
and 
\be
(a) + \mbox{Fig.~(\ref{Box1bis})} + (b) = \frac{1}{s}\left(\log^2\frac{s}{M^2} + \log^2\frac{u}{M^2} \right).
\ee
Reinserting factors of 2 and couplings, we find the full crossed contribution relative to the 
Born amplitude 
\be
c^{\rm ang}_{\rm crossed} = \frac{g_0^2}{32\pi^2}\left(\log^2\frac{s}{M^2} + \log^2\frac{u}{M^2}\right).
\ee 
Summing it with the direct term we obtain
\be
c^{\rm ang} = c^{\rm ang}_{\rm direct} + c^{\rm ang}_{\rm crossed} = 
-\frac{g_0^2}{32\pi^2}\left(\log^2\frac{t}{M^2} - \log^2\frac{u}{M^2}\right) 
\stackrel{NLO}{=}  
-\frac{g_0^2}{16\pi^2}\ \log\frac{s}{M_Z}\ \log\frac{t}{u}.
\ee 
where we neglect subleading terms $\sim \log^2(-t/s)$ or $\sim \log^2(-u/s)$ not growing with energy. 
Also, in the single logarithm, we have chosen arbitrarily the scale $M_Z$ which is not fixed at this order
in the logarithmic expansion. 

After the already discussed identification $g_0 = 2\ (g' Y/2)$, we find perfect agreement with the calculation in components. 
Expanding  $\int d^4\theta \ \Phi_1\Phi_2\overline\Phi_3\overline\Phi_4$ in component fields, we recover SSR for the angular Sudakov correction.
In addition, we have proved that SSR can be extended to the amplitudes for other four fermion or sfermion 
scattering processes, like for instance the fully scalar scattering 
$\widetilde f \widetilde f^*\to \widetilde f \widetilde f^*$. This is a pleasant result that 
can be checked by explicit component analysis~\cite{Fernand}.

As a technical remark, we notice that from the above expressions, it is straightforward to check that there are no 
box contributions of Yukawa type giving asymptotic Sudakov logarithms.
The diagrams proportional to $y_t^4$ with exchange of virtual $\Phi$ superparticles can be calculated
with the same techniques and all the terms growing with the energy cancel.

\section{Conclusions}
\label{Conclusions}

We have shown that superfield perturbation theory can be applied to the calculation 
of Sudakov electroweak corrections in the MSSM. The precise physical origin of these corrections 
permits to identify the supergraphs that produce logarithmically enhanced effects.
For simplicity, we have analyzed the neutral, essentially abelian, contributions.
The calculation is made easy by the physical nature of Sudakov corrections which are 
well definite, finite and gauge invariant.
With a single supergraph, we simultaneously calculate the correction for several possible external states.
In this framework, it is therefore quite straightforward to prove the existence of supersymmetric 
relations among the Sudakov corrections to various processes. 

It is clear that many extensions are possible, in particular to processes relevant to LHC. For instance, 
single top quark production via the parton processes $bu\to td$, $u\overline d\to t\overline b$, $bg\to tW^-$,
and $bg\to tH^-$ receives large SUSY electroweak corrections as discussed in~\cite{SudakovLHC}
at the Sudakov level.
The superfield approach can prove in such cases supersymmetric Sudakov relations with processes
of production of single stop squark that have similar large corrections. 

As a final comment, we emphasize the remarkable simplicity of the Yukawa correction in superfield language 
and the possibility of computing it in a gaugeless limit of the MSSM. Indeed, it should be possible
to evaluate the  phenomenologically important two loop contribution in a rather compact way,
at least in the Sudakov approximation, by combining recent results on multi-loop asymptotic expansions~\cite{DennerMultiLoop}
with superfield perturbation theory. Work is in progress on such topic.

We thank F. M. Renard for many discussions on radiative corrections in the MSSM, and
explicit calculations in the component approach.

\end{document}